\RequirePackage{fix-cm}
\documentclass{svjour3}
\usepackage{aas_macros}
\smartqed  % flush right qed marks, e.g. at end of proof
\usepackage{graphicx}
\usepackage[numbers]{natbib}
\usepackage{amsmath}	
\usepackage[T1]{fontenc}
\begin{document}

\title{Frequency shift in gravitational lensing by Kerr metric}

\titlerunning{Frequency shift by Kerr metric} % if too long for running head

\author{Samaneh Sarbaz \and Sohrab Rahvar}

\institute{Samaneh Sarbaz \at
              Department of Physics, Sharif University of Technology, Azadi ave. Tehran 11365-9161, Iran \\
              \email{samaneh\_sarbaz@physics.sharif.edu}           
           \and
           Sohrab Rahvar \at
              Department of Physics, Sharif University of Technology, Azadi ave. Tehran 11365-9161, Iran \\
              Research Center for High Energy physics, Sharif university of Technology, Tehran, Iran \\
              \email{rahvar@sharif.edu}
}

\date{Received: date / Accepted: date}
% The correct dates will be entered by the editor

\maketitle

\begin{abstract}
In this study, we calculate the relative frequency shift ($\Delta \nu / \nu$) caused by gravitational lensing resulting from the transverse motion of a spinning lens described by the Kerr metric \cite{Kerr:1963ud}. While the effect of transverse velocity on the frequency shift has previously been investigated using the Schwarzschild metric \cite{Rahvar_2020,10.1093/mnras/stae1938}, extending this to the Kerr metric provides a new window for measuring the transverse velocity of a lens or its spin parameter, a parameter often subject to degeneracy in gravitational microlensing observations. Recent microlensing observations have revealed black holes as viable lensing candidates \cite{2022ApJ...933...83S}. Since spinning black holes form naturally from stellar evolution, astrophysical black hole lenses are best described by the Kerr metric. Furthermore, recent observations of supermassive black holes at the centers of galaxies provide an ideal laboratory for testing the observational effects of the Kerr metric by measuring the frequency shifts of lensed sources. 

\end{abstract}

\section{Introduction}

Gravitational lensing stands as a cornerstone prediction of general relativity and has evolved into a remarkably powerful tool in modern astrophysics \citep{1936Sci....84..506E, 1992grle.book.....S}. The observable effects of lensing, such as image magnification and distortion, depend critically on the underlying spacetime geometry of the deflecting mass. While the Schwarzschild metric \citep{Schwarzschild:1916uq} serves as the foundational model for a non-rotating spherical mass, the reality is that many astrophysical objects possess intrinsic angular momentum. Therefore, the Kerr metric \citep{Kerr:1963ud}, characterized by the spin parameter $a={J}/{Mc}$ and the phenomenon of frame-dragging \citep{Penrose:1969pc,Chandrasekhar1983}, provides a more accurate and appropriate description for such rotating bodies. Extensive investigations into Kerr lensing have revealed that the spin of the lens introduces distinct, spin-dependent modifications to the resulting lensed images \cite{Hsieh_2021}.

Beyond these geometric effects, gravitational lensing, coupled with the transverse motion of the lens, can also induce a kinematic frequency shift in the light originating from the source. This phenomenon has been previously discussed and explored as a potential probe for non-rotating lenses \cite{Rahvar_2020,10.1093/mnras/stae1938}. However, the specific contribution of the lens's spin parameter to this frequency shift, particularly in the astrophysically relevant weak-field regime, has not been fully established. While microlensing is widely used to probe phenomena ranging from dark matter to exoplanets \cite{1993ApJ...404..441K, 1992ApJ...396..104G}, isolating the signature of the lens's spin remains a significant challenge.

This study aims to bridge this gap by systematically examining the influence of the Kerr spin parameter $a$ on the gravitational frequency shift and determining whether it creates a distinct, measurable weak-field signature. In particular, supermassive black holes (SMBHs) residing at the centers of galaxies present an ideal astrophysical laboratory for testing these effects. Due to their immense mass and potential for high spin, observing the frequency shifts of background sources lensed by such objects could provide a direct, measurable probe of the Kerr metric and the angular momentum of massive cosmic bodies.

This paper is organized as follows. In Section II, we introduce the theoretical framework of the Kerr metric. Section III details the calculation of the Shapiro time delay in this spacetime geometry. In Section IV, we derive the frequency shift induced by the Kerr metric, analyze the perturbative image positions, and present our numerical results for both stellar-mass and supermassive black holes. Section V provides a comparison of our approach with the second-order post-Minkowskian approximation. Finally, Section VI summarizes our conclusions.

\section{Kerr Metric}

The Kerr metric describes the spacetime geometry surrounding a rotating, uncharged, axially symmetric massive object. It is a vacuum solution to Einstein's field equations and generalizes the Schwarzschild solution by including angular momentum.

In Boyer--Lindquist coordinates \citep{Boyer:1967tg,Carroll:2004st} $(t, r, \theta, \phi)$, the space-time element is given by:
\begin{align} \label{eq:1}
  ds^2 =& -\left(1 - \frac{2 G M r}{\Sigma c^2}\right)c^2\,dt^2
- \frac{4 G M a r \sin^2\theta}{\Sigma c}\,dt\,d\phi \\ \notag
+& \frac{\Sigma}{\Delta}\,dr^2
+ \Sigma\,d\theta^2
+ \left(r^2 + a^2 + \frac{2 G M a^2 r \sin^2\theta}{\Sigma c^2}\right)\sin^2\theta\,d\phi^2,  
\end{align}

where
\[
\Sigma = r^2 + a^2 \cos^2\theta,
\quad
\Delta = r^2 - \frac{2 G M r}{c^2} + a^2.
\]

For the Kerr metric, we have the two event horizons for $\Delta=0$, i.e., $r_{\pm} = \frac{G M}{c^2} \pm \sqrt{\left(\frac{G M}{c^2}\right)^2 - a^2}$.

When the spin parameter $a$ vanishes, the Kerr metric reduces smoothly to the Schwarzschild metric with the known Schwarzschild radius of $r_s = {2 G M}/{c^2}$. In astrophysics, the Kerr solution is essential for modeling accretion disks, relativistic jets, and gravitational lensing in the vicinity of rotating black holes.

According to the standard definitions of the Kerr metric, the spin parameter $a$ can take values satisfying
\begin{equation}
0 \leq a \leq \frac{G M}{c^2},
\end{equation}
where the upper bound corresponds to an \emph{extremal Kerr black hole}, for which the event horizon and the ergosphere merge in the limiting case $a = G M / c^2$~\citep{Chandrasekhar1983}.

Crucially for the study of photon trajectories, the Kerr metric possesses a non-zero off-diagonal component, $g_{t\phi}$. This term is responsible for the frame-dragging (or Lense-Thirring) effect .

For the observable gravitational lensing phenomena targeted in this study occur at distances much larger than the gravitational radius of the lens, it is highly useful to evaluate the metric in the weak-field limit ($r \gg G M / c^2$ and $r \gg a$). Expanding the metric coefficients to leading order in $G$, the relevant components for calculating the photon transit time reduce to:
\begin{align}
    g_{tt} &\approx -\left(1 - \frac{2GM}{rc^2}\right), \\
    g_{rr} &\approx 1 + \frac{2GM}{rc^2}, \\
    g_{t\phi} &\approx -\frac{2GMa}{rc^3} \sin^2\theta.
\end{align}
This linearized form isolates the lowest-order contributions of the lens mass $M$ and the spin $a$, providing the necessary foundation to compute the Shapiro time delay and the subsequent kinematic frequency shift in the sections that follow.

\section{Shapiro Time Delay}

In the context of gravitational lensing, the total time delay experienced by a photon can be decomposed into two parts\citep{Narayan:1996ba}:
\begin{equation}
T(\vec{\theta}) = t_{\mathrm{geo}} + t_{\mathrm{grav}},
\end{equation}
where \(t_{\mathrm{geo}}\) represents the geometric delay due to the increased path length, and \(t_{\mathrm{grav}}\) corresponds to the Shapiro (gravitational) delay caused by the potential well of the lens and $\vec{\theta}$ is the angular position of the image from the lensing in two-dimension.

 The deflection angle $\vec{\alpha}$ is given by the gradient of the Fermat potential lensing $\psi$ \citep{1985A&A...143..413S,1986ApJ...310..568B,2015IJMPD..2430020R}. Following Fig.~\ref{fig:1}, the relation between the image position $\vec{\theta}$ and the source position $\vec{\beta}$ can be written as the lens equation relates the source position \(\vec{\beta}\) and the image position \(\vec{\theta}\) through the deflection angle \(\vec{\alpha}\):
\begin{equation}\label{eq:lenseq}
\vec{\beta} = \vec{\theta} - \vec{\alpha}(\vec{\theta}),
\qquad
\vec{\alpha}(\vec{\theta}) =\vec{ \nabla_{\!\theta} }\psi(\vec{\theta}),
\end{equation}
where \(\psi(\vec{\theta})\) is the two-dimensional lensing potential.

On the other hand, the time delay function can thus be written as
\begin{equation}
\tau(\vec{\theta}) = Q \left[ \frac{1}{2} |{{\vec{\theta} - \vec{\beta}}}|^2 - \psi(\vec{\theta}) \right],
\end{equation}
with the factor of
\begin{equation}
Q = \frac{D_L D_S}{c D_{LS}}\,
\end{equation}
where \(D_L, D_S,\) and \(D_{LS}\) denote the distances between observer–lens, observer–source, and lens–source, respectively. Varying $\tau$ with respect to $\vec{\theta}$ and equating it to zero we get the lens equation as  $\vec{\theta} - \vec{\beta} = \vec{\nabla}_\theta \psi$. So the arrival time when we have the image would be:
\begin{equation}
\tau(\vec{\theta}) = Q \left[ \frac{1}{2} (\nabla_\theta\psi)^2 - \psi(\vec{\theta}) \right],
\end{equation}
In the case of having lens moves with transverse to our line of sight, the arrival time as a function of $\theta$ can change. Now we are aiming to calculate the time derivative of photons arrival time,  i.e. $d\tau/dt$ which is
\begin{equation}\label{eq:301}
\dot{\tau}
= \vec{\nabla_{\!\theta}} \tau \cdot \dot{\vec{\theta}}
=Q \Big[\mathbf{H}_\psi - \mathbf{I}\,\Big] \vec\nabla_\theta\psi\cdot\dot{\vec{\theta}},
\end{equation} 
where we use the lensing equation and have the expression of 
$$\nabla_{\theta} |\vec{\theta} - \vec{\beta}|^2 = 2 \nabla_{\theta} \psi \cdot\nabla_{\theta} ( \nabla_{\theta}  \psi)
$$ 
and the Hessian matrix is
\begin{equation} \label{eq:hessian}
\mathbf{H}_\psi({\theta})
\;=\;
\nabla_{\!\theta}(\nabla_{\!\theta} \psi)
=
\begin{pmatrix}
\displaystyle
\frac{\partial^2 \psi}{\partial \theta_x^2}
&
\displaystyle
\frac{\partial^2 \psi}{\partial \theta_x \partial \theta_y}
\\[1.0em]
\displaystyle
\frac{\partial^2 \psi}{\partial \theta_y \partial \theta_x}
&
\displaystyle
\frac{\partial^2 \psi}{\partial \theta_y^2}
\end{pmatrix}.
\end{equation}
This matrix describes how the deflection field varies locally across the lens plane .
Comparing equation (\ref{eq:301}) with that results from the Schwartzchild metric, we infer that the effect of the geometric term appears only as a multiplicative factor of  $(\mathbf{H}_\psi - \mathbf{I})$  in the gravitational time delay, i.e.  $Q\vec\nabla_\theta\psi\cdot\dot{\vec{\theta}}$.

 Allowing \(\vec{\beta}\) and \(\vec{\theta}\) to depend on time, from equation (\ref{eq:lenseq}) we have :
 \begin{equation}
\dot{\vec{\theta}} - \dot{\vec{\beta}}
= \frac{d}{dt}\big(\nabla_{{\theta}}\psi(\vec{\theta})\big).
\end{equation}
and we rewrite this equation as,
\begin{equation}
\dot{\vec{\theta}} - \dot{\vec{\beta}}
= \mathbf{H}_\psi(\vec{\theta}) \, \dot{\vec{\theta}}.
\end{equation}
or 
\begin{equation}
\dot{\vec{\theta}}
= \big(\mathbf{I} - \mathbf{H}_\psi(\vec{\theta})\big)^{-1} \dot{\vec{\beta}}.
\end{equation}

Substituting in (\ref{eq:301}) results in a simplifies form for the time derivative of the arrival time as  

\begin{equation}\label{eq:302}
\dot{\tau}
=Q  \vec\nabla _\theta\psi\cdot\dot{\vec{\beta}},
\end{equation} 
where $\dot{\vec{\beta}}$ is the angular velocity of the source star. We note that in this simplified equation, the time derivative of arrival time is given in terms of the angular velocity of the source star.

\section{Frequency shift by Kerr metric}
The frequency shift in gravitational microlensing for the Schwartzchild metric has been introduced in \cite{Rahvar_2020}\cite{10.1093/mnras/stae1938}. The frequency shift in the gravitational lensing means that if we take two images for a single lens, the transverse velocity of the lens causes the change of gravitational potential of the lens at the position of images. This results in a different time derivative as in equation (\ref{eq:302}) for the arrival time from two paths to the observer. So we expect to have a relative frequency shift during the lensing where with high resolution spectrographs this effect could be observable.

Here we intend to consider the Kerr metric for lensing and derive the relative frequency shift for this gravitational lensing.
The objective is to calculate the relative frequency shift, namely $\Delta \nu / \nu$ for each image as well as its deviation from the Schwarzschild frequency shift.
In our calculations, we assume that the source is located at 
$-\infty$ along the x-axis, while the observer is situated at $+\infty$ along the same axis. The gravitational lens, which is a rotating black hole, is positioned at $x=0$.
The spin axis of the lens (i.e., the spinning black hole) is aligned with the z-axis. 
Calculation for any arbitrary orientation of the spin is given in appendix \ref{a}.

 Figure (\ref{fig:1})  illustrates the light emit from the source is deflected by the lens along two different paths and reaches the observer, resulting in two distinct images of the source as perceived by the observer. These two images exhibit a time delay relative to each other due to the different trajectories they follow in the lens metric. A detailed discussion for the Kerr metric is provided in \cite{Hsieh_2021}.
 
In the following, we aim to investigate the contribution of the spinning lens to the frequency shift. 
In particular, we examine how the angular momentum of the lens modifies the time delay and, consequently, the observed frequency shift of the images. The Kerr geometry, describing the spacetime around a rotating black hole of mass $M$ and angular momentum $a$.

 \begin{figure}
     \centering
     \includegraphics[width=1\linewidth]{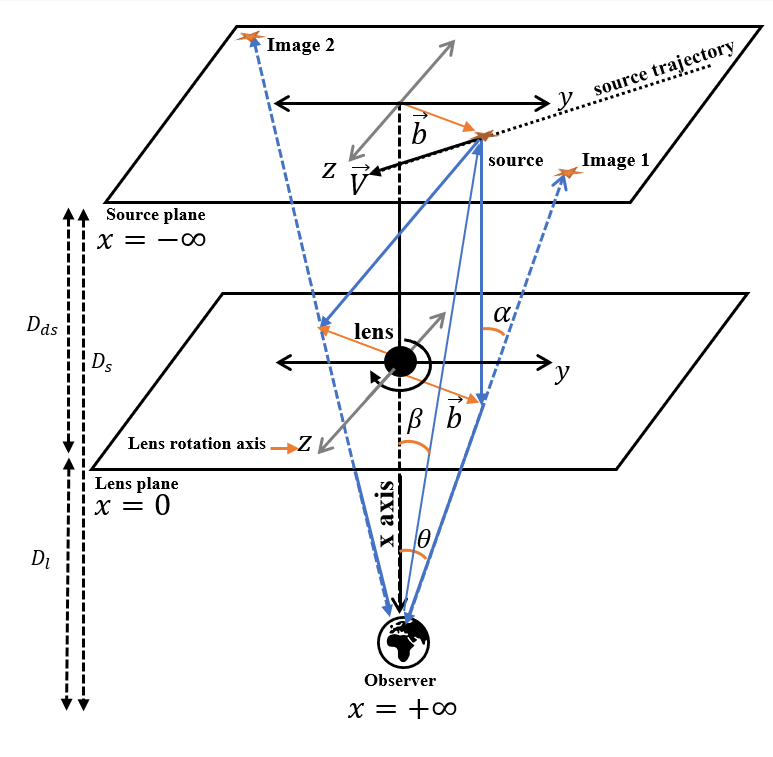}
     \caption{ Schematic illustration of gravitational lensing by a rotating black hole. Light emitted by the source at $-\infty$ along the X-axis is deflected by the lens located at $X=0$, whose spin axis is aligned with the Z-axis.The diagram shows an arbitrary point along the source's trajectory, with a relative transverse velocity with respect to the lens, given by $V^2 = V_z^2 + V_y^2.$ The impact parameter is denoted by \(b\), where $b^2 = z^2 + y^2$}
     \label{fig:1}
 \end{figure}

A prominent representation of the Kerr metric, first introduced in Kerr's seminal paper \cite{Kerr:1963ud} and further explored in a subsequent conference proceeding, is formulated using a pseudo-Cartesian coordinate system ($t, x, y, z$). This specific form, known as the Kerr-Schild metric \citep{visser2008kerrspacetimebriefintroduction} derived from equation (\ref{eq:1}) is suitable for our study and is given by
\begin{equation}\label{eq:12}
 ds^2 = \eta_{\mu\nu} dx^\mu dx^\nu + A \,(l_\mu dx^\mu)^2,   
\end{equation}
where
\[
A = \frac{2GM r^3}{c^2(r^4 + a^2 z^2)},
\]
and
\[
l_\mu dx^\mu = c\,dt + \frac{rx + ay}{r^2 + a^2}\,dx + \frac{ry - ax}{r^2 + a^2}\,dy + \frac{z}{r}\,dz. 
\]

Here, the term $\eta_{\mu\nu} dx^{\mu} dx^{\nu}$ corresponds to the standard Minkowski line element ($ds^2 = -c^2dt^2 + dx^2 + dy^2 + dz^2$). In particular, the radial coordinate $r$ is not an independent variable but instead serves as a function implicitly defined by the lens plane coordinates ($x, y, z$). This dependency is governed by the relation:
\begin{equation}
x^2 + y^2 + z^2 = r^2 + a^2 \left( 1 - \frac{z^2}{r^2} \right).
\end{equation}

This implicit definition of $r$ is a defining characteristic of the Kerr-Schild coordinate system.

For null geodesics ($ds^2=0$) restricted to $dy=dz=0$, equation (\ref{eq:12}) reduces to
\begin{equation}\label{eq:13}
0 = -c^2 dt^2 + dx^2 + A\,(c\,dt + B\,dx)^2, 
\end{equation}
where $B = \frac{rx+ay}{r^2+a^2}$ . Equation (\ref{eq:13})
yielding the quadratic equation
\begin{equation} \label{eq:14}
    c^2(A-1) dt^2 + 2 A c B\, dt\,dx + (1+AB^2)\,dx^2 = 0~ ,
\end{equation}
where solution is expressed as
\begin{equation}\label{eq:15}
    dt(r) = \frac{-A B \pm \sqrt{(A B)^2 - (A-1)(1+AB^2)}}{c (A-1)} dx~ .
\end{equation}
To compute the arrival time taken by the photon to travel along the path from the source at $-\infty$ to the observer at $+\infty$ along the x-axis, the integration is \begin{align}\label{eq:16}
   T = \int ^{+\infty} _{-\infty}dx \frac{-A B \pm \sqrt{(A B)^2 - (A-1)(1+AB^2)}}{c (A-1)}. 
\end{align}
{Since our objective is to calculate the frequency shift, it is preferable to first evaluate 
$\frac{\Delta\nu}{\nu}=-\frac{dT}{dt}$
before attempting to solve the integral. We know \(T=T(x,y,z;r)\), where \(r=r(x,y,z)\), so we apply the chain rule
\begin{equation}\label{eq:17}
    \frac{dT}{dt} = \frac{\partial T}{\partial x}\frac{dx}{dt}
+\frac{\partial T}{\partial y}\frac{dy}{dt}+
\frac{\partial T}{\partial z}\frac{dz}{dt}
\end{equation}

 $T$ is not an explicit function of time, implying that its partial derivative with respect to $t$ is zero.
By introducing lens velocity components $V_x=dx/dt ,V_y ={dy}/{dt}, V_z ={dz}/{dt}$ and $V_{\rm lens}=- V_{\rm source}$ where the observer is taken as the reference ,we will have
\begin{equation}\label{eq:def}
\frac{dT}{dt}
=\Big(\frac{\partial T}{\partial x}
+\frac{\partial T}{\partial r}\frac{\partial r}{\partial x}\Big)V_x+\Big(\frac{\partial T}{\partial y}
+\frac{\partial T}{\partial r}\frac{\partial r}{\partial y}\Big)V_y
+\Big(\frac{\partial T}{\partial z}
+\frac{\partial T}{\partial r}\frac{\partial r}{\partial z}\Big)V_z.
\end{equation}
}
In other words:
\begin{equation}\label{eq:18}
 \frac{\Delta\nu}{\nu}=-\frac{dT}{dt} = -\nabla T \cdot V_{\rm lens}
=\nabla T \cdot V_{\rm source}
\end{equation}

We substitute in equation (\ref{eq:302}), replacing the angular derivative with respect to the spatial derivatives results in 
\begin{equation}\label{eq:freqshift_tot}
     \frac{\Delta\nu}{\nu} =  \frac{D_l}{D_s}\, \nabla T . \, \vec{V}_{\text{\rm source}}
\end{equation}

To analyze the temporal behavior of the photons during the lensing event, we need to evaluate the partial derivatives of the total travel time $T$ with respect to the spatial coordinates $(x, y, z)$ and the radial distance $r$. This requires differentiating the integrand $I$ within the time integral under the integral sign. Since the spacetime geometry is governed by the Kerr metric, the functions $A$ and $B$ inherently depend on these spatial coordinates, making the explicit differentiation highly non-trivial. 

By applying the chain and quotient rules, we can analytically determine the exact derivatives of the integrand. This mathematical framework allows us to systematically account for the frame-dragging effects caused by the black hole's spin, as well as the non-linear geometric corrections. Furthermore, the detailed coordinate transformations mapping the angular lensing positions to physical coordinates on the lens plane are explicitly analyzed. To maintain the logical flow and readability of the primary text, the step-by-step differentiation, the explicit algebraic expressions for each partial derivative, and the comprehensive derivation of the Kerr lensing image positions are detailed in \textbf{Appendix~\ref{aa}}. 

Once these derived components are integrated along the photon trajectory using appropriate trigonometric substitutions, they are directly substituted into the total travel time framework and the corresponding frequency shift master equations to determine the final observable frequency shift. Crucially, to ensure that the solution remains physically valid and respects causality, we evaluate the system in the asymptotic Minkowski limit where gravity vanishes. This boundary condition uniquely isolates the physically realistic branch of the solution, establishing a robust foundation for the final lensing equations.

\subsection{Numerical Analysis and Results}

In this section, we provide a detailed numerical analysis by adopting physically motivated parameters. We consider a supermassive Kerr black hole with a mass of $M = 10^7 M_{\odot}$ as the gravitational lens, situated at a distance of $D_L = 1 \text{ Mpc}$ from the observer. The source is modeled as a distant quasar located at $D_S = 1 \text{ Gpc}$ with a transverse velocity of $v_s = 1000 \text{ km/s}$. To accurately highlight the contribution of the \textit{spin parameter} to the magnification, the impact parameter is set to $b = 10^{-5}$, which consistently satisfies the condition $\beta \gg \epsilon$ (see equation \ref{eq:AA17}). 

\begin{figure}
    \centering
    \includegraphics[width=1\linewidth]{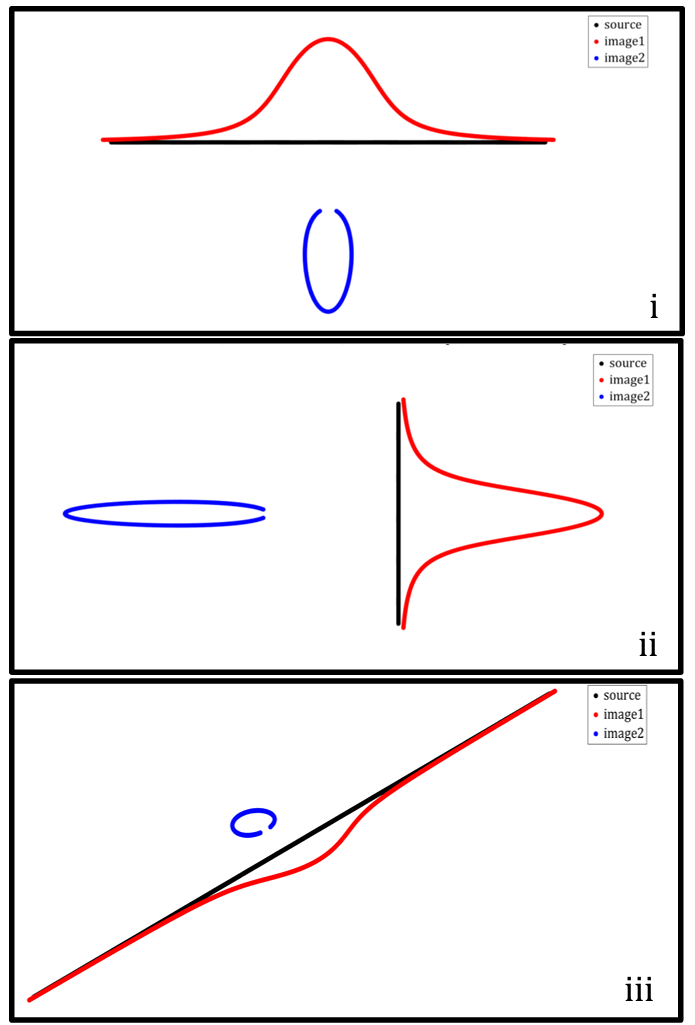}
    \caption{Schematic representation of the gravitational lensing geometry and the source's projected trajectories in the lens plane. The central black hole is characterized by its \textit{spin parameter} $a$. The three considered scenarios for the source's motion are depicted: (1) motion along the $y$-axis, (2) motion along the $z$-axis, and (3) motion in an arbitrary direction. The impact parameter $b=10^{-5}$ is chosen to ensure the validity of the weak-field approximation $\beta \gg \epsilon$. The labels Image 1 and Image 2 represent the two images formed in the non-relativistic limit.}
    \label{fig:panel}
\end{figure}

The investigation is categorized into three primary scenarios based on the projected motion of the source in the lens plane, as depicted in Figure \ref{fig:panel}: (i) motion along the $y$-axis, (ii) motion along the $z$-axis, and (iii) motion in an arbitrary direction.It should be emphasized that while Image 1 and Image 2 are clearly separated and occupies distinct spatial regions, the influence of the \textit{spin parameter} on the specific trajectory of each individual image is negligible. Specifically, when viewing the lens system without extreme magnification, the paths traced by Image 1 for $a=0$, $a=0.5M$, and $a=0.9M$ effectively coincide, appearing as a single trajectory. The same behavior holds for Image 2. This lack of spatial sensitivity to the black hole's spin underscores the importance of the frequency shift analysis, as it provides a more robust channel for detecting the \textit{spin parameter} than the geometric positioning of the images alone.Under the non-relativistic approximation, the Kerr lens generates two primary images, hereafter referred to as \textit{Image 1} and \textit{Image 2}. As the source moves away from the lens, \textit{Image 1} approaches the source's position while \textit{Image 2} moves toward the lens center. 

The numerical results for these three scenarios are illustrated in Figures 2, 3, and 4, each consisting of three vertical panels. The \textbf{top panels} provide a detailed view of the frequency shift during the interval of closest approach to the lens at $t=0$, where the influence of the \textit{spin parameter} is most prominent. The \textbf{middle panels} depict the differential frequency shift between the two formed images, highlighting the relative variation between Image 1 and Image 2 across the source trajectory. Finally, the \textbf{bottom panels} illustrate the deviation of the frequency shift in the Kerr metric from the Schwarzschild case ($\frac{\Delta \nu_{Kerr}}{\nu} - \frac{\Delta \nu_{Sch}}{\nu}$). This layout allows for a comprehensive comparison of how the \textit{spin parameter} modulates the frequency characteristics of the system, both relative to the individual images and to the non-rotating limit.

\begin{figure}
    \centering
\includegraphics[width=1\linewidth]{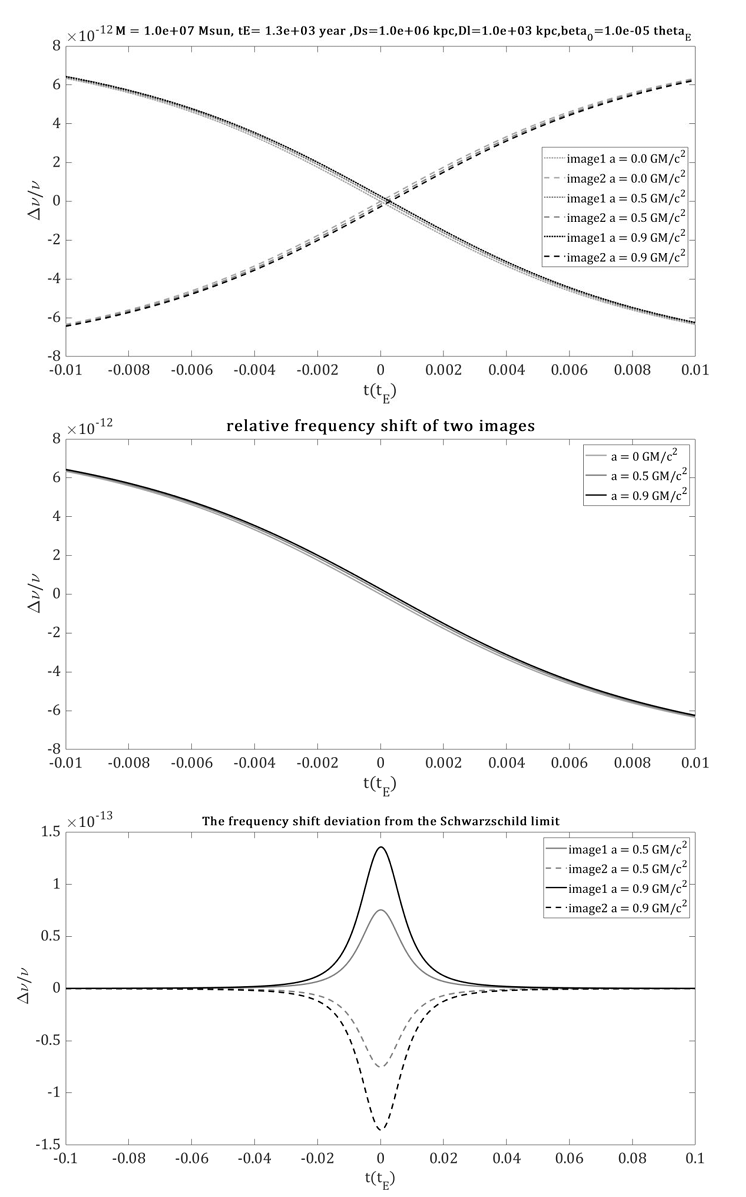}
    \caption{Frequency shift analysis for Case I (motion along the $y$-axis). \textbf{Top panel:} Close-up view of the frequency shift for both images near the point of closest approach ($t=0$). \textbf{Middle panel:} The difference in frequency shift between Image 1 and Image 2 ($\Delta \nu_1/\nu - \Delta \nu_2/\nu$). \textbf{Bottom panel:} The residual frequency shift relative to the Schwarzschild baseline ($\Delta \nu_{Kerr}/\nu - \Delta \nu_{Sch}/\nu$), illustrating the specific contribution of the \textit{spin parameter}.}
    \label{fig:9}
\end{figure}
\begin{figure}
    \centering
    \includegraphics[width=1\linewidth]{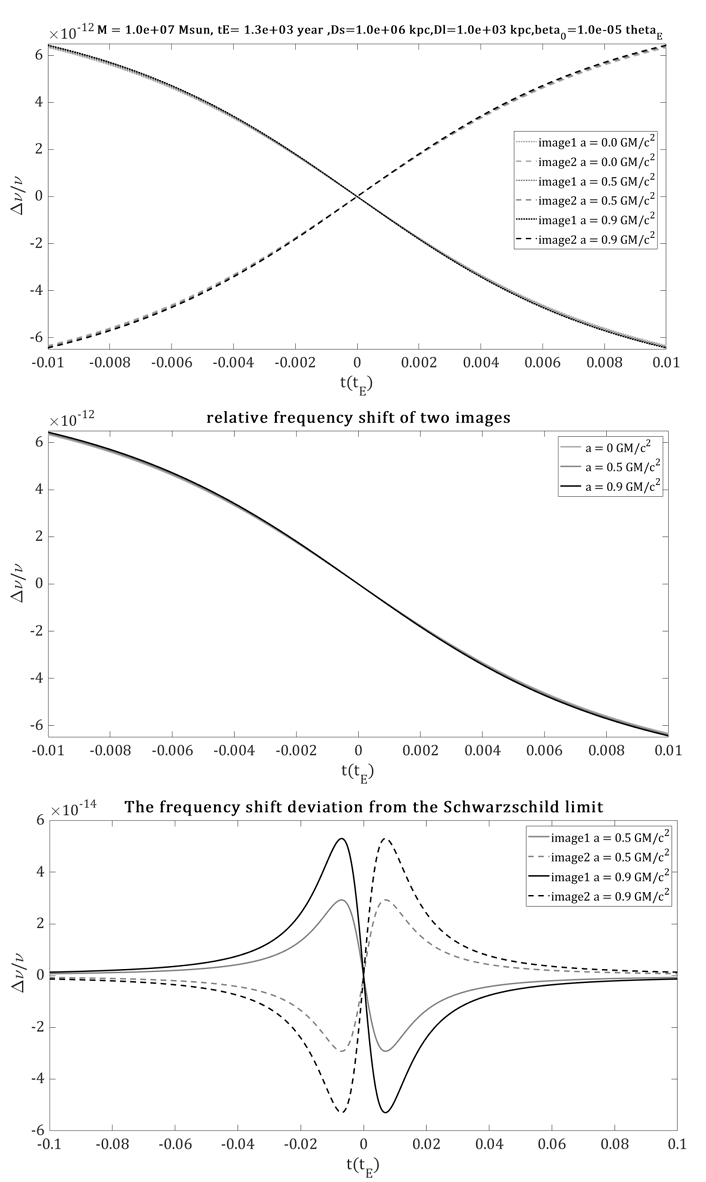}
    \caption{Frequency shift analysis for Case II, where the projected motion of the source is aligned with the $z$-axis. \textbf{Top panel:} Close-up view of the frequency shift for both Image 1 and Image 2 near the point of closest approach ($t=0$). \textbf{Middle panel:} The differential frequency shift between the two images ($\Delta \nu_1/\nu - \Delta \nu_2/\nu$), showing the relative variation during the source's transit. \textbf{Bottom panel:} The residual frequency shift relative to the Schwarzschild baseline ($\Delta \nu_{Kerr}/\nu - \Delta \nu_{Sch}/\nu$), highlighting the specific modulation caused by the \textit{spin parameter} $a$ in this geometric configuration.}
    \label{fig:10}
\end{figure}

\begin{figure}
    \centering
    \includegraphics[width=1\linewidth]{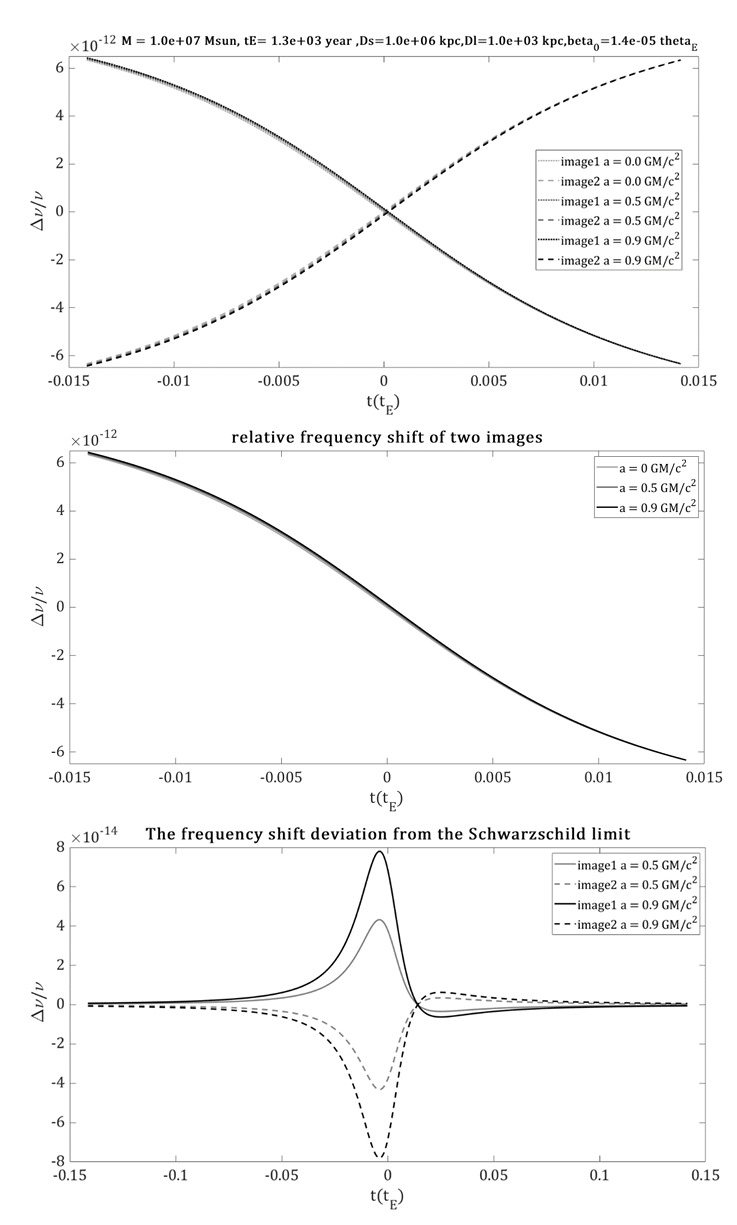}
    \caption{Frequency shift analysis for Case III, where the projected motion of the source follows an arbitrary direction in the lens plane. \textbf{Top panel:} Close-up view of the frequency shift for Image 1 and Image 2 during the interval of closest approach ($t=0$). \textbf{Middle panel:} The differential frequency shift between the two images ($\Delta \nu_1/\nu - \Delta \nu_2/\nu$), illustrating the relative shift evolution for an arbitrary trajectory. \textbf{Bottom panel:} The residual frequency shift relative to the Schwarzschild limit ($\Delta \nu_{Kerr}/\nu - \Delta \nu_{Sch}/\nu$), demonstrating how the \textit{spin parameter} $a$ modulates the signal when the symmetry of the motion is broken.}
    \label{fig:3}
\end{figure}

From the inspection of the figure \ref{fig:10}, it can be inferred that the spin parameter can enhance the effect of the frequency shift. To provide a clearer understanding, we also present an additional plot illustrating the frequency shifts of the two images produced in the Kerr gravitational lensing configuration. In addition, the relative frequency shift between these two images is plotted, offering a direct comparison of their differential behavior.(see figure \ref{fig:3})

The numerical results align with the theoretical predictions derived from the frequency shift relation $\Delta \nu \propto \frac{4GM}{b} \hat{b} \cdot \vec{V}$, as previously discussed in \cite{10.1093/mnras/287.4.833}. Our analysis confirms that for the negative parity image (Image 2), which approaches the lens as the source recedes, the frequency shift diverges toward larger values. However, as this image approaches the strong-field regime, it nears the photon capture limit. As established by \cite{Chandrasekhar1983}, the critical impact parameter for photon capture is given by $b_{crit} = 3\sqrt{3}R_s$. This threshold marks the boundary beyond which light rays are gravitationally captured. To maintain the validity of the weak-field approximation while exploring high-sensitivity proximity, we focused on a negative parity image situated at $10 R_s$. While this image exhibits a significant frequency shift, its magnification decreases substantially, rendering it increasingly faint. 

\begin{figure}
    \centering
    \includegraphics[width=1\linewidth]{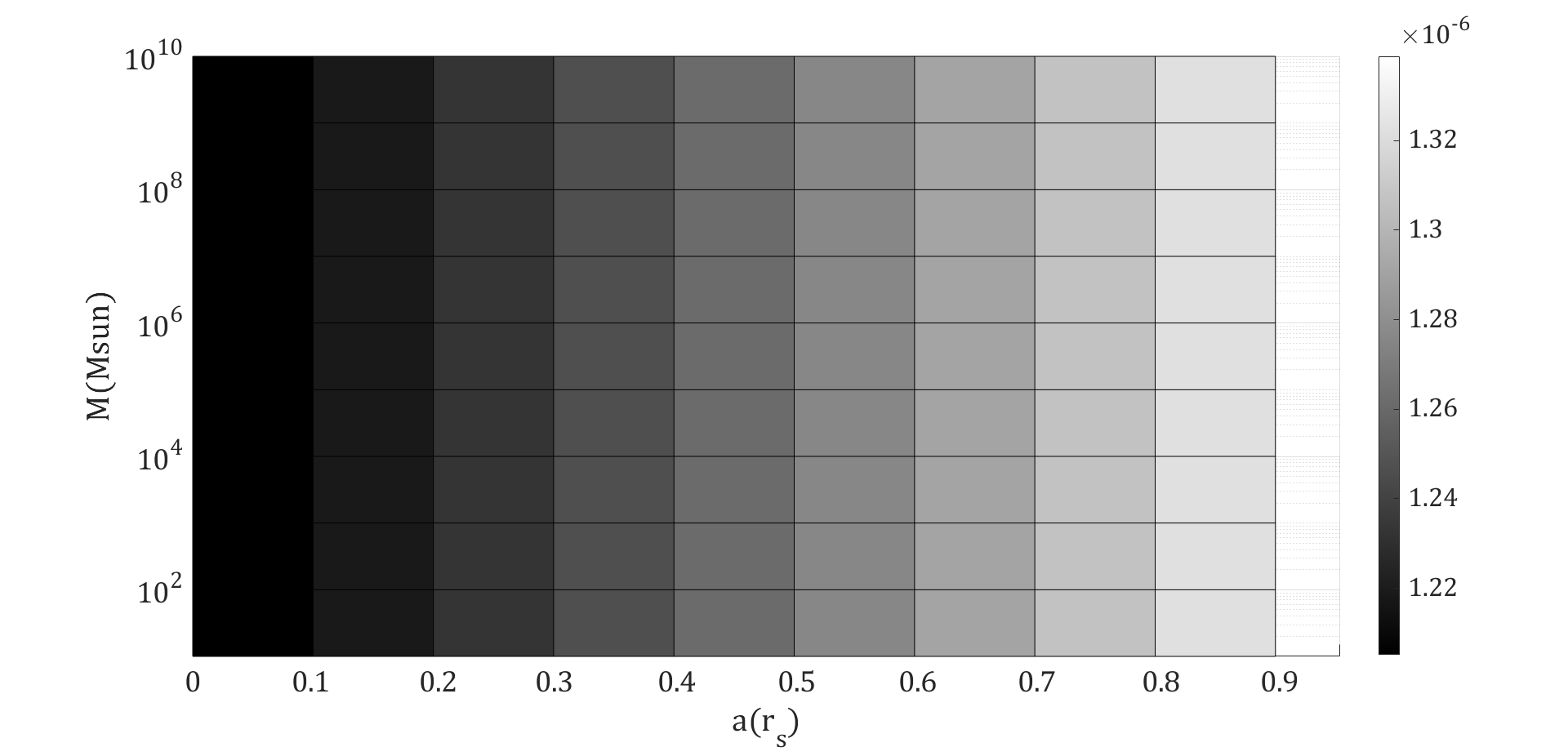}
    \caption{Distribution of the frequency shift for the negative parity image at a close proximity of $10 R_s$ from the lens. The color bar represents the frequency shift on a linear scale, ranging from the minimum values (black) to the maximum values (light gray). At this distance, the contours illustrate the high sensitivity of the frequency shift to the \textit{spin parameter}, showing that the frequency shift modulation is primarily governed by the black hole's spin rather than its mass.}
    \label{fig:222}
\end{figure}

\begin{figure}
    \centering
    \includegraphics[width=1\linewidth]{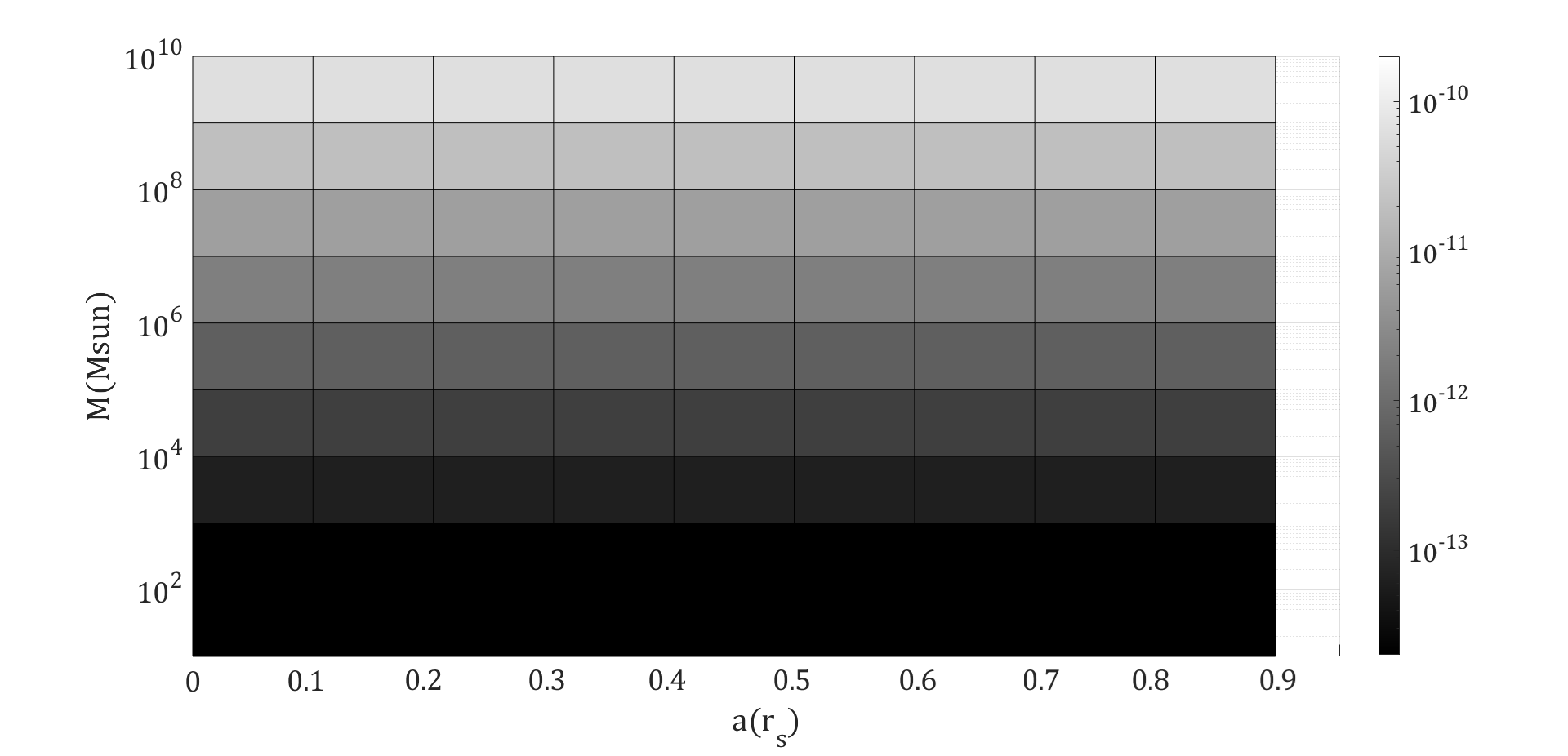}
    \caption{Frequency shift contours for the negative parity image formed at a larger distance of $0.5 R_E$. Due to the significant variation in the order of magnitude of the frequency shift—driven predominantly by the lens mass in this regime—a logarithmic color bar is employed. The scale transitions from the lowest values in black to the highest values in light gray. This visualization highlights how the gravitational potential of the mass dominates the frequency shift characteristics at distances further from the relativistic influence of the \textit{spin parameter}. }
    \label{fig:333}
\end{figure}

A remarkable finding, illustrated in Figure \ref{fig:222}, is that the frequency shift at this proximity is predominantly sensitive to the \textit{spin parameter} rather than the lens mass. Our simulations indicate that a stellar-mass black hole ($M \approx M_{\odot}$) generates a frequency shift comparable to that of a supermassive black hole at this specific distance, provided the \textit{spin parameter} is identical. In contrast, for images formed at larger distances (e.g., $0.5 R_E$), Figure \ref{fig:333} demonstrates that the lens mass dictates the order of magnitude of the shift, necessitating a logarithmic color scale for effective visualization. 

These observations suggest a novel dual-diagnostic approach for characterizing black holes through the decoupling of mass and spin effects. To practically implement this method, it is imperative to identify the primary source and correlate it with its corresponding lensed images through high-resolution spectroscopic or photometric observations. Once the images are matched, the \textit{spin parameter} can be directly inferred from the near-lens images frequency shift, while the mass can be determined with high precision from images located further from the lens (e.g., at $0.5 R_E$). This framework provides a robust tool for testing the properties of Kerr spacetime.

To investigate the mass-dependence of the integral and simplify the analytical derivations, we restrict our analysis to the equatorial plane by setting $z=0$. This physically motivated assumption significantly streamlines the metric derivatives. For instance, the vertical gradient of the potential in Eq.~\eqref{eq:AA3} vanishes ($\partial A/\partial z = 0$), which consequently eliminates the $z$-directed variation of the integrand presented in Eq.~\eqref{eq:AA9}. Furthermore, the radial gradient in Eq.~\eqref{eq:AA10} reduces to the standard planar form, specifically $\partial r/\partial y = y/r$. As a direct mathematical consequence, the total out-of-plane contribution to the variation, given by $\frac{\partial I}{\partial z} + \frac{\partial I}{\partial r} \frac{\partial r}{\partial z}$, identically vanishes.
To formally extract the mass scale from the integration, we use the Schwarzschild radius, $R_s = 2GM/c^2$, as the natural characteristic length of the system. We parameterize the impact parameter and the Kerr spin parameter in terms of this scale, such that $y = \eta R_s$ (where $\eta$ is a dimensionless constant, set to $10$ in our numerical evaluation) and $a = a_* R_s$, where $a_*$ is the dimensionless spin parameter. Evaluating the fundamental metric function $A$ under these normalizations yields $A = R_s/r$. Consequently, both $A$ and the function $B$ become purely geometric, dimensionless quantities that are fundamentally independent of the black hole mass $M$.
To evaluate the integral along the photon trajectory, we apply the substitution $x = w \tan u$, where $w = \sqrt{y^2-a^2}$. Under this transformation, the path differential becomes $dx = R_s \sqrt{\eta^2-a_*^2} \sec^2 u \, du$, demonstrating that $dx \propto R_s^1$. 
We now examine the dimensional scaling of the integrand components. The explicit spatial derivative $\partial I/\partial y$ given by Eq.~\eqref{eq:AA5} contains the rational term $a/(r^2+a^2)$, which scales as $R_s^{-1}$. Similarly, for the radial derivative $\partial I/\partial r$ defined in Eq.~\eqref{eq:AA11}, the evaluation of Eq.~\eqref{eq:AA13} at $z=0$ reduces to $A' = -R_s/r^2$, implying that $A'$ and consequently $\partial I/\partial r$ also scale as $R_s^{-1}$. The geometric term $\partial r/\partial y = y/r$ remains a dimensionless ratio of length scales.
When constructing the total variation along the line of sight in the equatorial plane, we evaluate the integral:
\begin{equation}
    \int_{-\infty}^{+\infty} \left( \frac{\partial I}{\partial y} + \frac{\partial I}{\partial r} \frac{\partial r}{\partial y} \right) dx
\end{equation}
The $R_s^{-1}$ proportionality of the partial derivatives within the parentheses perfectly cancels the $R_s$ proportionality of the path differential $dx$. Thus, the final evaluated integral is strictly a function of the dimensionless spin parameter $a_*$. 
It is crucial to note that relaxing the equatorial assumption (i.e., for $z \neq 0$) does not alter this fundamental scaling behavior. By employing an analogous dimensional argument for the non-vanishing $\partial I/\partial z$ and $\partial r/\partial z$ terms, one can explicitly show that the full generalized integral remains entirely independent of the black hole mass $M$. This confirms the mass-independence of our calculations and fully justifies the generality of the obtained results.


\subsection{Spin Parameter Estimation via Dual Symmetric Negative-Parity Images from Distinct Sources}

In this section, we propose a novel spectroscopic methodology leveraging a unique configuration to isolate and measure the black hole's spin parameter. In an innovative approach, we investigate the scenario featuring two negative-parity images originating from two distinct background source stars. These images are assumed to be captured on opposite sides of the Kerr black hole within the lens plane. Provided that advanced astronomical instruments can spatially resolve and distinguish these high-magnification images, specific narrow emission or absorption spectral lines—such as the Hydrogen ($\text{H}\alpha$) or Helium lines—can be systematically analyzed to extract their precise gravitational and kinematical frequency shifts. 

By comparing the frequency shifts of these corresponding, stable spectral lines for two images situated at well-defined, small impact parameters relative to the compact object, the intrinsic spin parameter of the black hole can be successfully constrained.For illustrative evaluation, we numerically simulate two images originating from two quasar sources located at a distance of 1 Gpc. These sources are gravitationally lensed by a fixed Kerr black hole situated at a distance of 1 Mpc. For simplicity, we assume that the two resulting images are positioned symmetrically on opposite sides of the lens exactly along the $y$-axis, at a close radial distance of $10 R_s$. Furthermore, a constant relative velocity with fixed components $V_y$ and $V_z$ is assigned to both configurations. It is important to emphasize that these parameters are chosen purely for demonstrative purposes; all such assumptions can be arbitrarily modified within our numerical simulation framework to yield correspondingly updated results. Finally, we calculate the differential relative frequency shift by subtracting the shift of one image from the other.

\begin{figure}
    \centering
    \includegraphics[width=1\linewidth]{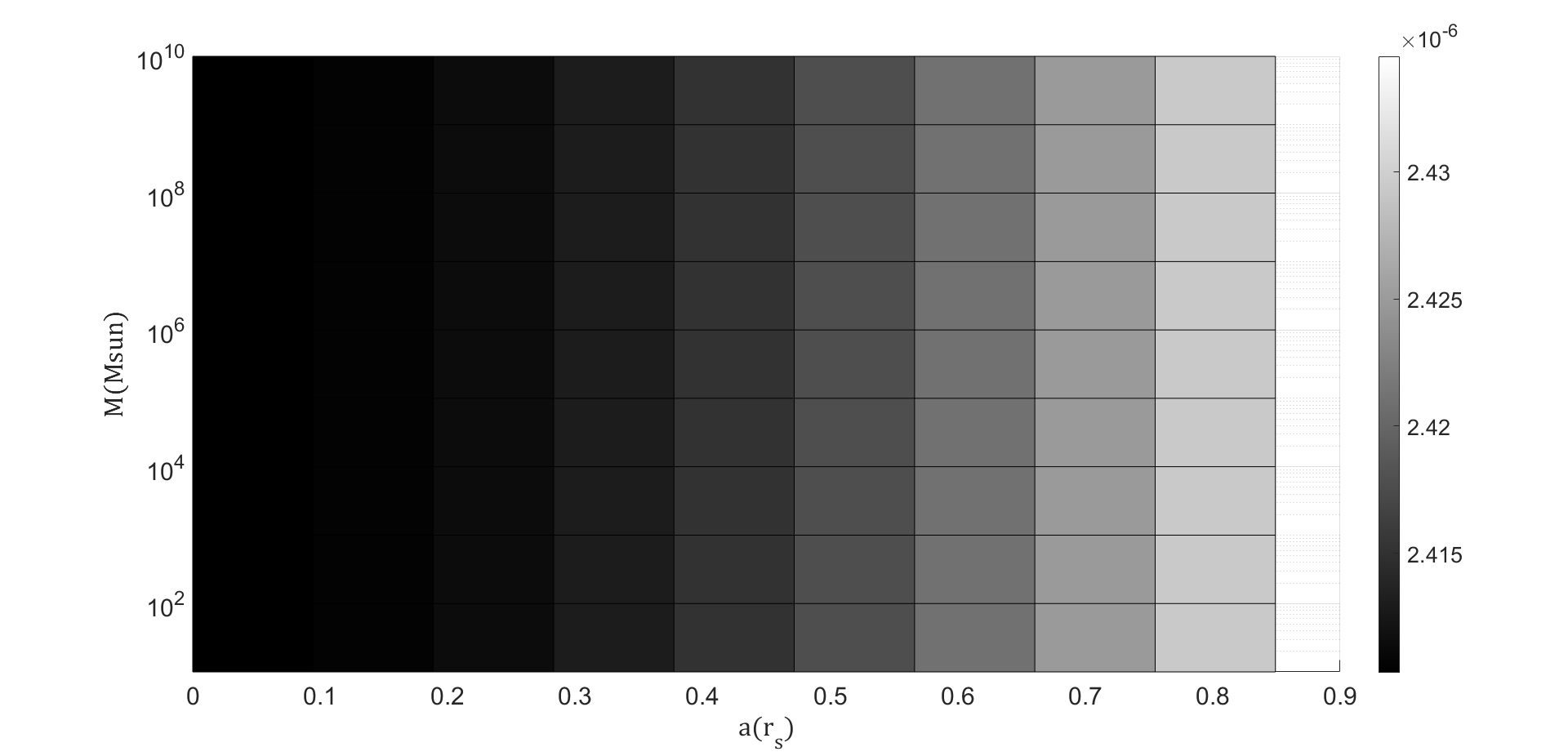}
    \caption{The differential relative frequency shift between two symmetric negative-parity images (located symmetrically on the $y$-axis of the lens plane where the impact parameter is $10R_s$) as a function of the black hole's spin parameter. As theoretically expected, for a non-rotating black hole (where the spin parameter is zero), the spacetime geometry is perfectly spherically symmetric, and consequently, the differential frequency shift between the two corresponding images vanishes exactly to zero.}
    \label{fig:1010}
\end{figure}

Our numerical simulations yield remarkable insights; the resulting differential relative frequency shift exhibits a strong, explicit dependence on the black hole's spin parameter, while being entirely independent of the lens mass $M$ (figure \ref{fig:1010}). This mass-independent feature provides a powerful mechanism to break the parameter degeneracy typically encountered in strong-field gravitational lensing. 

\section{Comparison with the Second-Order Post-Minkowskian Approximation}

To verify the analytical integrity of our results, we compare our formalism with the findings of . \cite{Jiang_2024}, who investigated the frequency shift of light in the gravitational field of a rotating mass. Conducted within the framework of Einstein’s general relativity, their study employs the second-order post-Minkowskian $(2PM)$ approximation to analyze light propagation in the Kerr spacetime. The authors consider scenarios involving the relative motion of both the source and the observer, presenting two complementary methodologies: one based on the proper time intervals between successive light waves, and another comparing photon energies at emission and detection. 

To establish a direct comparison, we adapt the final exact formula from their work, which formulates the observed-to-emitted frequency ratio in the asymptotic limit where both the source and observer are located at infinity. We reformulate their expression into a relative frequency shift under our specific kinematic configuration. Specifically, we assume a non-zero transverse velocity for the observer with respect to the lens while keeping the source at rest. This effectively reconstructs our physical configuration, wherein the lens moves relative to a stationary source and an observer.

Starting from the general asymptotic equation in \cite{Jiang_2024}:
\begin{equation}\label{eq:jiang_ratio}
\frac{\nu_o}{\nu_s}
= \frac{\gamma_o}{\gamma_s}\,
\frac{1-\mathbf{c}_o\!\cdot\!\mathbf{v}_o}{1-\mathbf{c}_s\!\cdot\!\mathbf{v}_s},
\end{equation}
where the photon direction vector at the source is unperturbed,
\begin{equation}
    \mathbf{c}_s=\mathbf{n}_s,
\end{equation}
and the photon direction vector at the observer, including the $2PM$ deflection, is given by:
\begin{align}\label{eq:jiang_co}
\mathbf{c}_o
=\left(1-\frac{8m^2}{b^{2}}\right)\mathbf{n}_s
&-\left[\frac{4m}{b^{2}}+\frac{15\pi m^{2}}{4b^{3}}
+\frac{4\,(\mathbf{n}_s\times\mathbf{b})\!\cdot\!\mathbf{J}}{b^{4}}\right]\mathbf{b}\notag \\
&-\frac{4\,\mathbf{b}\!\cdot\!\mathbf{J}}{b^{4}}\,
(\mathbf{n}_s\times\mathbf{b}).
\end{align}
The symbols are defined in the harmonic coordinate system (using geometric units $G=c=1$) as follows:
\begin{description}
  \item[$\nu_s,\,\nu_o$] Photon frequency at the source and the observer.
  \item[$\gamma_s,\,\gamma_o$] Lorentz factors of the source and observer ($\gamma=1/\sqrt{1-\mathbf{v}^2}$).
  \item[$\mathbf{n}_s$] Unperturbed propagation direction of the photon in flat spacetime.
  \item[$\mathbf{b}$] Impact parameter vector, orthogonal to $\mathbf{n}_s$, with magnitude $b=|\mathbf{b}|$.
  \item[$m$] Lens mass parameter ($m = GM/c^2$).
  \item[$\mathbf{J}$] Spin angular momentum of the lens ($\mathbf{J} = G\mathbf{J}/c^3$).
\end{description}

To align this general framework with our default conditions, we impose the following kinematic constraints:
\begin{equation}
\mathbf{n}_s=\hat{\mathbf{x}},\quad
\mathbf{b}=b_y\hat{\mathbf{y}}+b_z\hat{\mathbf{z}}, \quad
\mathbf{v}_o=v_o\hat{\mathbf{y}},\quad
\mathbf{J}=J\hat{\mathbf{z}},\quad
\gamma_o \approx \gamma_s \approx 1.
\end{equation}
Furthermore, we assume that source velocity is $\mathbf{v}_s=0$. Consequently, the term in the denominator of Eq. \ref{eq:jiang_ratio} involving the dot product of the photon's propagation direction and the source's velocity identically vanishes ($1-\mathbf{c}_s\!\cdot\!\mathbf{v}_s = 1$). 

Under these non-relativistic velocity limits for the observer ($\gamma_o \approx 1$), the fractional frequency shift $\Delta\nu/\nu \approx (\nu_o - \nu_s)/\nu_s$ can be extracted by substituting $\mathbf{c}_o$ into Eq. \ref{eq:jiang_ratio}. The relative frequency shift takes the following form:
\begin{equation}
\frac{\Delta\nu}{\nu}
=  v_o\!\left[
\Big(\frac{4m}{b^2}+\frac{15\pi m^2}{4b^3}+\frac{4b_y J}{b^4}\Big)b_y
-\frac{4b_z^2 J}{b^4}\right].
\end{equation}

It is crucial to discuss the fundamental agreement and the intrinsic theoretical deviations between our Kerr-Schild approach and the harmonic-gauge approach utilized by Jiang et al. 

Through explicit analytical calculations, it can be demonstrated that both frameworks yield identical linear contributions at the first-order approximation $\mathcal{O}(m)$. This exact correspondence at the 1PM level confirms the consistency of the Newtonian limit and verifies that the fundamental gravitational frequency shift mechanisms are fully preserved in both methodologies.

However, slight structural deviations emerge in the higher-order terms, specifically at the second-order approximation $\mathcal{O}(m^2)$. This is a well-documented characteristic of general relativity, fundamentally rooted in gauge dependence and the choice of coordinate systems. Jiang et al. employ harmonic coordinates, which are optimized for asymptotic flatness but possess inherent coordinate singularities near the event horizon. In contrast, our calculations are deeply embedded within the Kerr-Schild coordinate system. The Kerr-Schild metric is inherently horizon-penetrating and strictly regular across the horizon, leading to a mathematically distinct coupling between the spatial and temporal components of the metric ($g_{0i} \neq 0$) from the outset. 

Consequently, the precise definition of the spatial radial distance $r$ and the transverse impact parameter $b$ differs fundamentally between the harmonic and Kerr-Schild manifolds at $\mathcal{O}(m^2)$. Therefore, the minor discrepancies in the second-order coefficients are not computational discrepancies but rather coordinate artifacts reflecting the differing geometric mappings of the spacetime.

\section{Conclusions}

In this work, we have comprehensively investigated the gravitational frequency shift of light within the framework of gravitational lensing in the Kerr spacetime. By analyzing both the geometric and gravitational components of the total time delay, we established that the geometric contribution can be elegantly incorporated by scaling the gravitational time delay with a straightforward geometric factor. Utilizing the horizon-penetrating and singularity-free Kerr-Schild metric under the null geodesic condition ($ds^2 = 0$), we derived the total analytical gravitational time delay for a photon propagating along the unperturbed trajectory (the $x$-axis) from past null infinity ($-\infty$) to future null infinity ($+\infty$) via the Shapiro time delay methodology.

By taking the total time derivative of this integrated time delay, we successfully extracted the fractional frequency shift experienced by the received photon relative to its emission from the source. Our analytical results explicitly demonstrate that this frequency shift is tightly coupled to the fundamental parameters of the lensing system, including the lens mass ($m$), its spin angular momentum ($a$), the lens-observer distance, and the relative spatial configuration of the source.

Numerical evaluations and analytical analysis tailored to specific astrophysical scenarios revealed that variations in the lens spin exert a pronounced and observationally distinguishable influence on the profile of the frequency shift. Leveraging these findings, we proposed two novel methods for extracting the spin of a rotating compact object. First, we demonstrated that by analyzing the frequency shift profile of a negative-parity image located in the strong-field regime (sufficiently close to the lens) and cross-referencing it with a source positioned in the far-field zone, the lens spin can be uniquely determined. Second, we extended this approach to spectroscopic observations, showing that comparing the relative frequency shift of a specific emission or absorption line across two negative-parity images in close proximity to the lens offers a robust and practical technique for measuring the black hole spin.

Finally, to validate our theoretical model, we conducted a rigorous cross-comparison with the recent second-order post-Minkowskian (2PM) formulations derived by Jiang et al. \cite{Jiang_2024}. Our analysis confirmed that both frameworks converge on identical order-of-magnitude estimates and share a flawless agreement at the first-order linear approximation $\mathcal{O}(m)$, thereby verifying the accuracy of our weak-field baseline and the underlying gravitational redshift mechanisms. The subtle structural discrepancies emerging at the second post-Minkowskian order $\mathcal{O}(m^2)$ were shown to be pure coordinate artifacts rather than physical contradictions. These differences naturally arise from the geometric mapping variations between the harmonic gauge employed by Jiang et al. and the off-diagonal, regularized Kerr-Schild coordinate system utilized in our work. Ultimately, our Kerr-Schild framework provides a highly stable, singularity-free analytical platform well-suited for characterizing strong-gravity lensing phenomena and probing the intrinsic properties of rotating compact objects.

%\begin{acknowledgements}
%If you'd like to thank anyone, place your comments here
%and remove the percent signs.
%\end{acknowledgements}

% BibTeX users please use one of
%\bibliographystyle{plainnat}      % author-year citations
%\bibliographystyle{spmpsci}      % mathematics and physical sciences
\bibliographystyle{spphys}       % APS-like style for physics
\bibliography{example}   % name your BibTeX data base

\appendix
\section{Analytical Derivatives and Image Coordinate Transformations}\label{aa}
\subsection{Derivation of the Integrand Formulations and Frequency Shift Substitutions}
To compute the partial derivatives of the total travel time $T$, we must first differentiate the time element integrand $I$ with respect to the spatial components. We define the core integrand as:
\begin{equation}
I=\frac{N}{c(A-1)},
\end{equation}
where the numerator $N$ takes the quadratic form:
\begin{equation}
N=-AB \pm R,  \quad R=\sqrt{(AB)^2-(A-1)(1+AB^2)}=\sqrt{AB^2-(A-1)}.
\end{equation}
The total derivatives of travel time required for the primary lensing framework are then expressed through the following integral formulations:
\begin{align}\label{eq:1020_app}
    \frac{\partial T}{\partial x} = \int_{-\infty}^{+\infty} \frac{\partial I}{\partial x} \, dx, \quad &\frac{\partial T}{\partial y} = \int_{-\infty}^{+\infty} \frac{\partial I}{\partial y} \, dx, \quad  \frac{\partial T}{\partial z} = \int_{-\infty}^{+\infty} \frac{\partial I}{\partial z} \, dx,\notag \\& \frac{\partial T}{\partial r} = \int_{-\infty}^{+\infty} \frac{\partial I}{\partial r} \, dx.
\end{align}
At a fixed radial distance $r$, the partial derivatives of the metric functions $A$ and $B$ with respect to the Cartesian coordinates are given by:
\begin{equation}\label{eq:AA1}
\frac{\partial A}{\partial x} = 0,
\qquad
\frac{\partial B}{\partial x} = \frac{r}{\,r^{2}+a^{2}\,},
\end{equation}
\begin{equation}\label{eq:AA2}
\frac{\partial A}{\partial y}=0, \qquad
\frac{\partial B}{\partial y}=\frac{a}{r^2+a^2},
\end{equation}
\begin{equation}\label{eq:AA3}
\frac{\partial B}{\partial z}=0, \qquad 
\frac{\partial A}{\partial z}
=-\frac{4GM r^3 a^2 z}{c^2(r^4+a^2 z^2)^2}.
\end{equation}
Applying these components to the numerator derivative $\partial N/\partial y$ yields:
\begin{equation}\label{eq:AA4}
\frac{\partial N}{\partial y}
=A\,\frac{\partial B}{\partial y}\Big(-1\pm\frac{B}{R}\Big),
\end{equation}
which leads directly to the spatial derivative of the integrand with respect to $y$:
\begin{equation}\label{eq:AA5}
\frac{\partial I}{\partial y}
=\frac{A}{c(A-1)}\,
\frac{a}{r^2+a^2}\,
\Big(-1\pm\frac{B}{R}\Big).
\end{equation}
Similarly, differentiating $N$ with respect to the $x$-coordinate yields:
\begin{equation}\label{eq:AA6}
\frac{\partial N}{\partial x}
  = A \frac{r}{r^{2}+a^{2}}
    \left( -1 \pm \frac{B}{R} \right),
\end{equation}
which produces the explicit expression for $\partial I/\partial x$:
\begin{equation}
\frac{\partial I}{\partial x}
  = \frac{A}{c(A-1)}
    \frac{r}{r^{2}+a^{2}}
    \left( -1 \pm \frac{B}{R} \right).
\end{equation}
For the $z$-directed variation, utilizing the quotient rule gives:
\begin{equation}\label{eq:AA7}
\frac{\partial I}{\partial z}
=\frac{(A-1)\frac{\partial N}{\partial z}
 - N \frac{\partial A}{\partial z}}
      {c(A-1)^2}.
\end{equation}
Since $B$ does not explicitly depend on $z$, the derivative of the numerator simplifies to:
\begin{equation}\label{eq:AA8}
\frac{\partial N}{\partial z}
=\frac{\partial A}{\partial z}
\Big(-B\pm\frac{B^2-1}{2R}\Big).
\end{equation}
Combining these expressions provides the final compact form for the $z$-derivative of the integrand:
\begin{equation}\label{eq:AA9}
\frac{\partial I}{\partial z}
=\frac{\partial A/\partial z}{c(A-1)^2}
\left[
(A-1)\Big(-B\pm\frac{B^2-1}{2R}\Big)
 -(-AB\pm R)
\right].
\end{equation}

To fully account for the explicit spatial dependence of $r$ on the coordinates $x, y,$ and $z$, we calculate the respective partial gradients:
\begin{equation}\label{eq:AA10}
\frac{\partial r}{\partial x}
=\frac{x\,r^{3}}{r^{4}+a^{2}z^{2}}
,\quad \frac{\partial r}{\partial y} = \frac{yr^3}{r^{4}+a^{2}z^{2}} , \quad \frac{\partial r}{\partial z} = \frac{zr(r^2 + a^2)}{r^{4}+a^{2}z^{2}}.
\end{equation}
The radial derivative $\partial I / \partial r$ can be organized as follows:
\begin{equation}\label{eq:AA11}
    \frac{\partial I}{\partial r} = \frac{N'(r) D(r) - N(r) D'(r)}{D^2(r)},
\end{equation}
where $D(r)=c(A-1)$ represents the denominator, and the corresponding radial derivatives $N'(r)$ and $D'(r)$ are defined as:
\begin{equation}\label{eq:AA12}
    N'(r) \;=\; -\bigl(A'B + A B'\bigr) 
\;\pm\; \frac{A'(B^2-1) + 2AB\,B'}{2\sqrt{(AB)^2 - (A-1)(1+AB^2)}},
\end{equation}
\begin{equation}
    D'(r) \;=\; cA'.
\end{equation}
Here, the basic radial derivatives of the metric profiles $A$ and $B$ are written as:
\begin{equation}\label{eq:AA13}
    A' \equiv \frac{dA}{dr}
= \frac{2GM}{c^{2}}\;\frac{-\,r^{6}+3a^{2}z^{2}r^{2}}{(r^{4}+a^{2}z^{2})^{2}},
\end{equation}
\begin{equation}\label{eq:AA14}
    B' \equiv \frac{dB}{dr}
=\frac{x(r^{2}+a^{2})-(rx+ay)\,2r}{(r^{2}+a^{2})^{2}}.
\end{equation}

To evaluate the partial derivatives of the total travel time $T$ required for Equation~\ref{eq:1020_app}, the integrations of these derived components are analytically performed by employing appropriate trigonometric substitutions from $-\infty$ to $+\infty$. Once these integrated components are successfully obtained, the resulting analytical derivatives of $T$ are directly substituted into the main frequency shift equation (Equation~\ref{eq:freqshift_tot} in the main text) to yield the final, explicit expression for the observable frequency shift.

To uniquely identify the physically valid branch of the quadratic solution for the time element, the system is solved in the asymptotic Minkowski limit ($A \to 0$). In this flat spacetime regime, the expression reduces to $dt = \mp dx / c$. Because the photon moves forward along the $x$-axis from the source to the observer ($dx > 0$), causality requires a positive advancement in coordinate time ($dt > 0$). This constraint means the negative branch of the radical must be exclusively chosen, recovering the standard free-space limit $dt = dx / c$ when gravitational effects are absent.

\subsection{Image Positions and Physical Coordinates}
To practically evaluate these frequency shift equations, the angular lensing coordinates must be mapped onto physical lengths in the lens plane. The spatial coordinates $(x, y)$ utilized in the differentiation framework represent the physical image positions, denoted by the vector $\boldsymbol{\xi}$. As the source star progresses along its trajectory over time, the corresponding physical image positions must be continuously updated for each instantaneous source location. These physical coordinates are converted from angular positions via:
\begin{equation}\label{eq:AA16}
    \boldsymbol{\xi} = R_E \boldsymbol{\theta}.
\end{equation}

Following the analytical framework for weak deflection gravitational lensing by a Kerr black hole \cite{Sereno_2006}, the normalized angular positions of the images $\boldsymbol{\theta}$ are expanded as a Taylor series up to the second order in terms of a small expansion parameter $\epsilon$. Consistent with defining the radial distances from the lens origin, this parameter is defined as:
\begin{equation}\label{eq:epsilon_def}
    \epsilon \equiv \frac{\theta_E}{4D},
\end{equation}
where $D \equiv r_s/(r_o+r_s)$, with $r_o$ and $r_s$ being the radial distances from the lens to the observer and the source, respectively. The perturbative expansion for the image positions is thus given by:
\begin{equation}\label{eq:AA17}
    \theta_i^{\pm} = \theta_{i(0)}^\pm + \theta_{i(1)}^\pm \epsilon + \theta_{i(2)}^\pm \epsilon^2 + \mathcal{O}(\epsilon^3),
\end{equation}
where the superscripts $(\pm)$ denote the primary (positive parity) and secondary (negative parity) images, and $i \in \{1,2\}$ indicates the spatial components in the sky plane. The zeroth-order terms $\theta_{i(0)}$ match the standard Schwarzschild lens geometry, while the first- and second-order terms introduce the frame-dragging effects of the black hole spin parameter $a$ and higher-order geometric corrections.

It is crucial to note that for this analytical formalism to remain valid, the source impact parameter $\beta$ must strictly satisfy the condition $\beta \gg \epsilon$. When the source approaches the optical axis such that $\beta \sim \mathcal{O}(\epsilon)$, the perturbative expansion mathematically breaks down. In this limit, the higher-order correction terms in the Taylor series dynamically diverge (scaling proportionally to $1/\beta$) and become comparable to the main zeroth-order term. This artificial crossing causes spurious divergences in the numerical profiles, which do not represent physical caustics but rather the limitations of the weak-field approximation.
For an observer situated in the equatorial plane ($\mu_0 = 0$), the specific components at each order are determined as follows. The zeroth-order terms depend entirely on the source position vector $\boldsymbol{\beta} = (\beta_1, \beta_2)$:

\begin{align}\label{eq:AA18}
    \theta_{1(0)}^{\pm} &= \frac{1}{2} \left( 1 \pm \sqrt{1 + \frac{4}{\beta^2}} \right) \beta_1, \\
    \theta_{2(0)}^{\pm} &= \frac{1}{2} \left( 1 \pm \sqrt{1 + \frac{4}{\beta^2}} \right) \beta_2,
\end{align}
where $\beta^2 = \beta_1^2 + \beta_2^2$ and the unperturbed angular magnitude is defined as $\theta_{(0)} = \sqrt{\theta_{1(0)}^2 + \theta_{2(0)}^2}$.

The first-order perturbations, which incorporate the dimensionless spin parameter $a_m = a/M$, are expressed as:
\begin{align}\label{eq:AA19}
    \theta_{1(1)} &= \frac{15\pi}{16(1+\theta_{(0)}^2)} \frac{\theta_{1(0)}}{\theta_{(0)}} + \frac{(1-\theta_{1(0)}^2+\theta_{2(0)}^2) a_m}{1-\theta_{(0)}^4}, \\
    \theta_{2(1)} &= \frac{15\pi}{16(1+\theta_{(0)}^2)} \frac{\theta_{2(0)}}{\theta_{(0)}} - \frac{2\theta_{1(0)}\theta_{2(0)} a_m}{1-\theta_{(0)}^4}.
\end{align}
The second-order perturbations include higher-order geometric corrections alongside quadratic spin effects ($a_m^2$):
\begin{align}\label{eq:AA20}
    \theta_{1(2)} &= \Theta_{Sch}^{(2)} \frac{\theta_{1(0)}}{\theta_{(0)}} + \frac{16D^2}{3}\theta_{1(0)}\theta_{2(0)}^2 + a_m \Psi_1 + a_m^2 \Omega_1, \\
    \theta_{2(2)} &= \Theta_{Sch}^{(2)} \frac{\theta_{2(0)}}{\theta_{(0)}} + \frac{16D^2}{3}\theta_{1(0)}^2\theta_{2(0)} + a_m \Psi_2 + a_m^2 \Omega_2,
\end{align}
where $D = D_{ls}/D_s$ represents the ratio of the lens-source distance to the observer-source distance. The shared second-order Schwarzschild geometric factor is given by:
\begin{align}\label{eq:AA21}
    \Theta_{Sch}^{(2)}& = -\frac{225\pi^2(2\theta_{(0)}^2+1)}{256\theta_{(0)}(\theta_{(0)}^2+1)^3} \notag \\
   & -8\left[\frac{3(\theta_{(0)}^4-\theta_{(0)}^2-1)D_l^2 - 3D_{ls}(\theta_{(0)}^4-\theta_{(0)}^2+3)D_l}{3(D_l+D_{ls})^2\theta_{(0)}(1+\theta_{(0)}^2)} \right. \notag\\
  &\quad \quad \quad \left.+ \frac{D_{ls}^2(2\theta_{(0)}^6-7\theta_{(0)}^4+6\theta_{(0)}^2-6)}{3(D_l+D_{ls})^2\theta_{(0)}(1+\theta_{(0)}^2)}\right].
\end{align}
\begin{align}\label{eq:AA22}
    \Psi_1 &= 5\pi\left[\frac{\theta_{(0)}^2(1+\theta_{(0)}^2)^2(1+4\theta_{(0)}^2)}{16\theta_{(0)}^3(1-\theta_{(0)}^2)(1+\theta_{(0)}^2)^3}\right.\notag \\ & \quad \quad \quad \ \left.-\frac{(12\theta_{(0)}^6+5\theta_{(0)}^4+4\theta_{(0)}^2-1)\theta_{1(0)}^2}{16\theta_{(0)}^3(1-\theta_{(0)}^2)(1+\theta_{(0)}^2)^3} \right]\\
    \Omega_1 &= -\theta_{1(0)} \left[ \frac{4(\theta_{(0)}^4+\theta_{(0)}^2+1)\theta_{2(0)}^2}{(1-\theta_{(0)}^2)^2(\theta_{(0)}^2+1)^3} - \frac{\theta_{(0)}^2}{(\theta_{(0)}^2+1)^3} \right]
\end{align}
and for the second spatial component, these auxiliary functions are written as:
\begin{align}\label{eq:AA23}
    \Psi_2 &= \frac{5\pi(-12\theta_{(0)}^6-5\theta_{(0)}^4-4\theta_{(0)}^2+1)\theta_{1(0)}\theta_{2(0)}}{16\theta_{(0)}^3(1-\theta_{(0)}^2)(\theta_{(0)}^2+1)^3}, \\
    \Omega_2 &= \frac{\theta_{2(0)} \left( \theta_{(0)}^2(3\theta_{(0)}^4+2\theta_{(0)}^2+3) - 4(\theta_{(0)}^4+\theta_{(0)}^2+1)\theta_{2(0)}^2 \right)}{(1-\theta_{(0)}^2)^2(\theta_{(0)}^2+1)^3}.
\end{align}

Because gravitational lensing alters the apparent surface area of the source, the final observable effect represents a combined manifestation of both the underlying kinematics and the specific geometry of the deflected light rays.
\section{Arbitrary direction for spin parameter}
\label{a}
From the Kerr–Schild with an arbitrary spin vector \cite{PhysRevD.103.084029} , \cite{combi2024binaryblackholemetric}:
\begin{equation}\label{eq:ks_def}
g_{ab}=\eta_{ab}+2H(\mathbf X),l_a l_b,
\end{equation}

where $\eta_{ab}$ is the Minkowski metric in Cartesian coordinates $X^a=(t,x,y,z)$ and the scalar field $H(\mathbf X)$ is :

\begin{equation}\label{eq:H_def}
 H(\mathbf X)=\frac{GM r^3}{c^2(r^4 + (a_i X^i)^2)}=\frac{M r^3}{r^4 + (a_x x + a_z z)^2},
\end{equation}

and the null covector $l_a$ fully determine the geometry where :

\begin{equation}\label{eq:l_general}
l_a dX^a = cdt + \frac{1}{r^2+a^2}\Big[r X_i- \epsilon_{ijk} a^j X^k + \frac{(a_i X^i)a_i}{r}\Big]dX^i 
\end{equation}

 Without loss of generality we choose the spatial spin vector
\begin{equation}\label{eq:spin_choice}
\mathbf a=(a_x,0,a_z),
\end{equation}

because any general spin vector may be rotated about the $z$-axis so that its $y$-component vanishes; the remaining two components $a_x,a_z$ parametrize the magnitude and orientation of the spin in the $x$--$z$ plane.

where
\begin{equation}\label{eq:chi_def}
\chi\equiv a_i X^i = a_x x + a_z z,
\end{equation}

and $r$ is the Boyer–Lindquist-like radial function implicitly defined by
\begin{equation}\label{eq:r_def}
r^2=\tfrac12(R^2-a^2)+\tfrac12\sqrt{(R^2-a^2)^2+4\chi^2}
\end{equation}
Where $R^2=x^2+y^2+z^2$.

Using equation (\ref{eq:ks_def}) $ds^2$ is:
\begin{equation}\label{eq:ds_full}
ds^2 = -c^2dt^2 + dx^2 + dy^2 + dz^2 + 2H(\mathbf X)\big(l_a dX^a\big)^2.
\end{equation}

As we said before with neglecting the effect of geometrical time delay in gravitational lensing frequency shift , we restrict attention to null path $ds^2=0$ with $dy=0,dz=0$. 
Under these conditions only $dt$ and $dx$ are nonzero. So equaton \ref{eq:l_general} will be:
\begin{equation}\label{eq:L_dx}
 l_a dX^a = cdt + A(\mathbf X),dx,
\end{equation}

where

\begin{align}\label{eq:A_def}
A(\mathbf X) &= \frac{1}{r^2+a^2}\Big[r x -\epsilon_{1jk}a^j X^k + \frac{\chi.a_x}{r}\Big] \notag \\
&= \frac{1}{r^2+a^2}\Big[r x + a_z y  + \frac{\chi.a_x}{r}\Big].
\end{align}

 by writing :
\begin{equation}
0 = -c^2dt^2 + dx^2 + 2H\big(dt + A.dx\big)^2.
\end{equation}
or
\begin{equation}\label{eq:quadratic_v}
(-1+2H) c^2dt^2 + 4H A \ cdt\ dx + (1+2H A^2)dx^2=0.
\end{equation}
we could solve (\ref{eq:quadratic_v}) for $dt$ :

\begin{equation}\label{eq:v_solution}
dt=\frac{-4H A \pm \sqrt{(4H A)^2 - 4(-1+2H)(1+2H A^2)}}{2c(-1+2H)} dx.
\end{equation}

So we can define $T=\int dt$ . According equation \ref{eq:def} its total time derivative m
ust be evaluated using the chain rule,
\begin{equation}
\frac{dT}{dt}
=
\frac{\partial T}{\partial x}\frac{dx}{dt}
+
\frac{\partial T}{\partial y}\frac{dy}{dt}
+
\frac{\partial T}{\partial z}\frac{dz}{dt}.
\end{equation}

Each spatial derivative decomposes as
\begin{equation}
\frac{\partial T}{\partial x^i}
=
\left.\frac{\partial T}{\partial x^i}\right|_{r}
+
\frac{\partial T}{\partial r}
\frac{\partial r}{\partial x^i},
\qquad
x^i\in\{x,y,z\},
\end{equation}
where the second term accounts for the implicit coordinate dependence of \(r\).

We define

\begin{equation}
\Delta =
\sqrt{\left(R^2-a^2\right)^2+4\chi^2},
\end{equation}

Calculate $\left.\frac{\partial T}{\partial x^i}\right|_{r}$ we need :
\begin{equation}
    \left.\frac{\partial T}{\partial x^i}\right|_{r}=\frac{\partial T}{\partial H}\left.\frac{\partial H}{\partial x^i}\right|_{r}+\frac{\partial T}{\partial A}\left.\frac{\partial A}{\partial x^i}\right|_{r}
\end{equation}

Keeping \(r\) fixed, the explicit spatial derivatives of \(H\) are
\begin{equation}
\left.\frac{\partial H}{\partial x}\right|_r
=
-\frac{2GM}{c^2}
\frac{r^3\chi a_x}{(r^4+\chi^2)^2},
\end{equation}
\begin{equation}
\left.\frac{\partial H}{\partial y}\right|_r = 0,
\end{equation}
\begin{equation}
\left.\frac{\partial H}{\partial z}\right|_r
=
-\frac{2GM}{c^2}
\frac{r^3\chi a_z}{(r^4+\chi^2)^2}.
\end{equation}

With \(r\) fixed, its explicit spatial derivatives of A are
\begin{equation}
\left.\frac{\partial A}{\partial x}\right|_r
=
\frac{r+\frac{a_x^2}{r}}{r^2+a^2},
\end{equation}
\begin{equation}
\left.\frac{\partial A}{\partial y}\right|_r
=
\frac{a_z}{r^2+a^2},
\end{equation}
\begin{equation}
\left.\frac{\partial A}{\partial z}\right|_r
=
\frac{a_x a_z}{r(r^2+a^2)}.
\end{equation}

The radial derivative of \(T\) is then
\begin{equation}
\frac{\partial T}{\partial r}
=
\frac{\partial T}{\partial H}\frac{\partial H}{\partial r}
+
\frac{\partial T}{\partial A}\frac{\partial A}{\partial r},
\end{equation}
with
\begin{equation}
\frac{\partial H}{\partial r}
=
\frac{GM}{c^2}
\frac{r^2(3\chi^2-r^4)}{(r^4+\chi^2)^2},
\end{equation}
and
\begin{equation}
\frac{\partial A}{\partial r}
=
\frac{x-\frac{a_x\chi}{r^2}}{r^2+a^2}
-
\frac{2r}{r^2+a^2}A.
\end{equation}

the derivatives of \(r\) with respect to the spatial coordinates are obtained by differentiating the implicit definition. One finds
\begin{equation}
\frac{\partial r}{\partial x}
=
\frac{1}{2r}
\left[
x
+
\frac{(R^2-a^2)x+2\chi a_x}{\Delta}
\right],
\end{equation}
\begin{equation}
\frac{\partial r}{\partial y}
=
\frac{y}{2r}
\left[
1+\frac{R^2-a^2}{\Delta}
\right],
\end{equation}
\begin{equation}
\frac{\partial r}{\partial z}
=
\frac{1}{2r}
\left[
z
+
\frac{(R^2-a^2)z+2\chi a_z}{\Delta}
\right].
\end{equation}

Substituting these expressions into the chain-rule expansion yields the final form of \(dT/dt\), corresponding to Eq.~(20), which consistently incorporates both explicit and implicit dependencies of the Kerr--Schild geometry with arbitrary spin orientation.

For the case where the lens is modeled as a supermassive black hole with a mass of $10^{10} M_{\odot}$, and the impact parameter is chosen as $b = 10^{-5} R_E$, various values of $a_x$ and $a_z$ were assigned. In this specific configuration, we have chosen the lens distance to be $16 \text{ Mpc}$ and the source distance to be $32 \text{ Mpc}$, representing an intergalactic space environment. Consequently, the lensing object is treated as an isolated supermassive black hole with a mass of $10^{10} M_{\odot}$.

In a generalized geometric configuration, we extended our analytical and numerical investigations to evaluate the influence of an arbitrarily oriented spin parameter on the gravitational frequency shift. By decomposing the black hole's spin vector into a longitudinal component parallel to the line of sight and transverse components residing within the lens plane, we sought to determine the individual contributions of these spatial projections.

Our results reveal a fundamental and striking decoupling: variations in the spi parameter along the line of sight have absolutely no effect on the magnitude or the profile of the frequency shift. The modulation of the frequency shift is exclusively governed by the transverse components of the spin vector projected onto the lens plane. This indicates that the longitudinal spin component remains optically inactive in the context of leading-order gravitational frequency shifts.

To rigorously verify the robustness of this longitudinal independence, we further expanded our kinematic model to include scenarios where the observer and the source possess non-zero velocity components along the line of sight. Remarkably, our analysis demonstrates that the structural independence of the frequency shift from the line of sight spin parameter strictly persists. Even in the presence of longitudinal relative motion, the coupling between the line-of-sight velocity and the line-of-sight spin component does not generate any additional frequency shift. This robust geometric feature significantly simplifies the parameter estimation process, as it allows observational data to constrain the transverse spin components without being contaminated by uncertainties in the longitudinal spin orientation.

\begin{figure}
    \centering
    \includegraphics[width=0.9\linewidth]{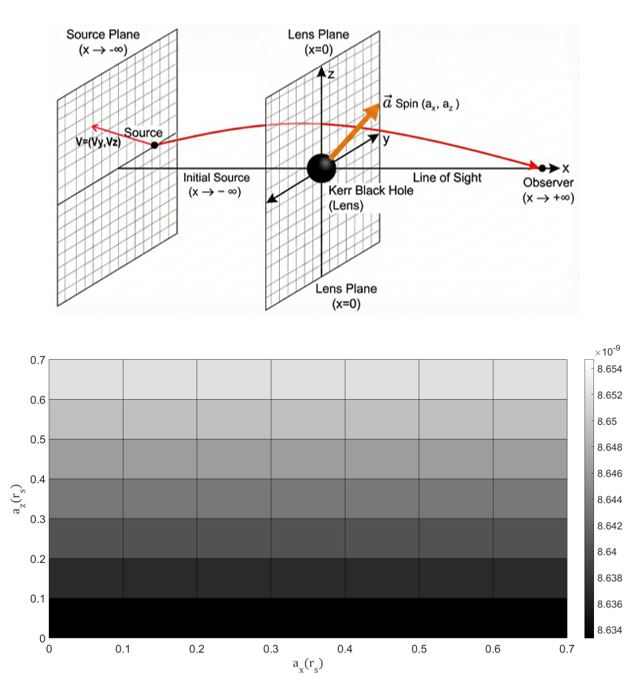}
    \caption{Frequency-shift diagrams for the configuration in which the source velocity lies entirely in the $y$--$z$ plane, i.e., parallel to the lens plane, and has no component along the line of sight. Figure consists of two panels.The top panel presents a schematic illustration of the overall geometry, showing the relative positions of the source, lens, and observer, as well as the orientation of the lens spin vector.The bottom panel displays the variation of the maximum frequency shift as a function of $a_z$ and $a_x$ }
    \label{fig:vyvz}
\end{figure}

\begin{figure}
    \centering
    \includegraphics[width=0.9\linewidth]{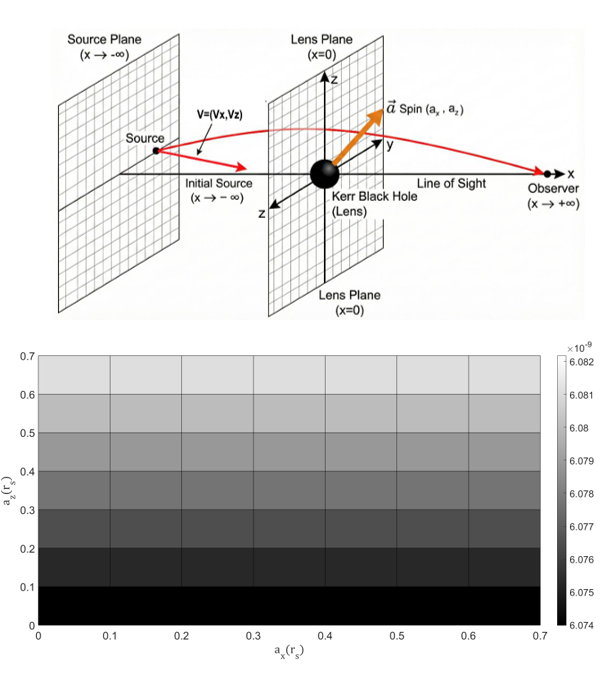}
    \caption{Frequency-shift diagrams for the configuration in which the source has a velocity component along the line of sight.Figure consists of two panels.The top panel presents a schematic illustration of the overall geometry, showing the relative positions of the source, lens, and observer, as well as the orientation of the lens spin vector.The bottom panel displays the variation of the maximum frequency shift as a function of $a_z$ and $a_x$
}
    \label{fig:vxvz}
\end{figure}

We present the frequency-shift diagrams for two distinct scenarios. 
In the first scenario, the source velocity lies entirely in the $y$--$z$ plane, i.e., it is parallel to the lens plane, and therefore has no component along the line of sight.( figure \ref{fig:vyvz})
In the second scenario, the source has a velocity component along the line of sight. (figure \ref{fig:vxvz})

Figures \ref{fig:vyvz} and \ref{fig:vxvz} summarize the main characteristics of these configurations. 

\end{document}